# Physics for Neuromorphic Computing

Danijela Markovic[1], Alice Mizrahi[1], Damien Querlioz[2], Julie Grollier[1]

[1] - Unité Mixte de Physique, CNRS, Thales, Univ. Paris-Sud, Université Paris-Saclay, 91767 Palaiseau, France

[2] - Université Paris-Saclay, CNRS, Centre de Nanosciences et de Nanotechnologies, 91120 Palaiseau, France.

**Neuromorphic computing takes inspiration from the brain to create energy efficient hardware for information processing, capable of highly sophisticated tasks. In this article, we make the case that building this new hardware necessitates reinventing electronics. We show that research in physics and material science will be key to create artificial nano-neurons and synapses, to connect them together in huge numbers, to organize them in complex systems, and to compute with them efficiently. We describe how some researchers choose to take inspiration from artificial intelligence to move forward in this direction, whereas others prefer taking inspiration from neuroscience, and we highlight recent striking results obtained with these two approaches. Finally, we discuss the challenges and perspectives in neuromorphic physics, which include developing the algorithms and the hardware hand in hand, making significant advances with small toy systems, as well as building large scale networks.**

# Why should we take inspiration from the brain?

Biological brains perform extremely complicated tasks because living beings need to solve complex problems for their survival. The smallest worm analyzes data, takes decisions and is able to move towards its goal. Brains furthermore compute with a remarkably low energy budget because living beings cannot use more energy than they can fetch and store. The human brain categorizes, predicts and creates with a power consumption only about 20 W. In contrast, our current digital computers are optimized for high-precision calculations, but consume an inordinate amount of energy when they run the type of cognitive tasks the brain excels at. For example, training a state-of-the art natural language processing model on a modern supercomputer consumes 1000 kW.h, which is the energy consumed by a human brain for the entirety of its tasks over a duration of six years (Box 1 provides supplementary information about this comparison). Brains are vastly different from human-made computing systems, both by the algorithms and their physical implementation.

***"Algorithms" of the brain.*** With its $10^{11}$ neurons interconnected through $10^{15}$ synapses, the human brain is a complex system that is dynamical, reconfigurable, and exhibits a wealth of fascinating phenomena for physicists, such as energy or entropy minimization[1,2], phase transitions and criticality[3], self-oscillations[4,5], chaos[6], synchronization[7,8], stochastic resonance[9] and many more. One of its most striking differences with conventional instruction-based algorithms is that it can learn from experience. From early on, physicists have participated in theoretical efforts to understand the algorithms of the brain, bringing contributions to both computer science and computational neuroscience. Statistical physics, non-linear dynamics, and complex systems theory have helped shed light on neural mechanisms that enable learning. In some cases, physicists have invented algorithms that rely on these physical principles for computing. Hopfield networks[1] and Boltzmann machines[10], which inherit from Ising spin systems are the most famous ones, but many other have been proposed, exploiting, in particular, non-linear dynamics for computing[11–13].

***Physical implementation.*** The brain as physical substrate of computing is also fundamentally different from our general purpose computers based on digital circuits[5,14]. It is based on biological entities such as synapses and neurons instead of memory blocks and transistors. It leverages the stochasticity of cells to function at very low energy, instead of relying on high precision circuits as in digital electronics (supplementary information about this topic is provided in Box 2). It uses both binary (single spikes) and analog (soma dynamics, synaptic weights) coding instead of just binary coding. As the brain generates its own internal temporal dynamics, it relies on asynchronous communication and avoids the high energy cost of the single clock used in computers for synchronous communication. The brain also embraces collective behaviors emerging in correlated devices for computing, instead of relying on independent circuits with well-defined functions. Finally, it exhibits extremely high plasticity due to its billions of synapses, compared to the limited re-configurability of digital electronics.

Our current processors rely on semiconductor physics combined with the Kirchhoff laws of electricity to implement efficiently Boolean logic. But as we will see, other physical principles and building blocks could be more adapted to implement neuromorphic chips inspired from the brain.

## Why are physics and material science essential to neuromorphic computing?

***Current electronics is not enough.*** In the brain, neurons, -- which can roughly be seen as carrying out the processing -- have a direct access to memory, supported by synapses. Current electronics, on the contrary, intrinsically separates memory and computing into distinct physical units, between which data must be carried back and forth. This "von Neuman bottleneck" is an issue for artificial intelligence algorithms which require reading considerable amounts of data at each step, performing complicated operations on this data, and then writing the results back to memory[15]. It slows down computing and considerably increases the energy consumption for learning and inference.

The general paradigm in neuromorphic computing is therefore to take inspiration from the topology of the brain to build circuits composed of physical neurons interconnected by physical synapses that implement memory in-situ, in a non-volatile way, thus drastically cutting the need to move data around the circuit and allowing huge gains in speed and energy efficiency. This is unfortunately complicated by using Complementary Metal Oxide Semiconductor (CMOS) technology alone. Dozens of transistors are needed to emulate each neuron, and additional external memories are required to implement synapses. CMOS-based artificial neurons and synapses are typically several micrometers wide[16], and the number of physical neurons and synapses that can be integrated in a CMOS chip is inherently limited by the chip area. This is problematic because the performance of neural networks increases with the number of neurons and synapses: typical image recognition algorithms today comprise millions of neurons and synapses in average. High numbers of neurons and synapses can be obtained by assembling chips together, but the whole system becomes bulky (the most complex versions of existing neuromorphic systems such as Spinnaker or Brainscales[17] can occupy several cubic meters) and a large part of the energy efficiency is lost in the interconnects. Nanodevices that can mimic important features of neurons and synapses at the nanoscale, such as non-linearity, memory and the ability to learn, are required to build low power chips comprising several millions of neurons and synapses.

Finally, it is difficult to achieve a high degree of interconnection between neurons using CMOS technology only. The brain features an average of 10,000 synapses per neuron. Such connectivity is impossible to reproduce with current electronics, as CMOS technology is mostly confined into two dimensions (2D), fan-out is limited, and it is difficult to efficiently and evenly supply energy to components in the circuit. On the contrary, the brain is tri-dimensional (3D), neuron axons and dendrites provide high fan-in/fan-out, and blood efficiently distributes supply energy to the whole system.

***What is needed (general considerations).*** The above considerations point out essential needs of neuromorphic computing that could be addressed by new materials and novel physical phenomena: nanoscale devices that imitate neurons and synapses with a low energy consumption and a high endurance, that are easy to address (read and write) in large networks, that provide signal gain and memory, that are tunable, active, dynamical, reconfigurable and multifunctional, that provide large fan-in / fan-out, large interconnectivity, can be self-assembled, can form 3D interconnects, are easily manufacturable in large quantities and, of course, at low cost.

## How can we imitate something that we do not understand yet?

How can these general goals be transformed in important contributions to the field of neuromorphic computing? As we do not have yet a precise model of how the brain is working, two main approaches are in current development. The first approach is to map conventional neural network algorithms – that are currently used in artificial intelligence – to dedicated physical systems, in order to achieve higher power efficiency when running them. The second approach is to reach beyond such algorithms, by taking inspiration from neuroscience to equip artificial neural network with additional features and dynamics, in hope to achieve more complex computing.

***Mapping Artificial Intelligence (AI) algorithms to physical systems.*** In recent years, considerable progress has been reported in the development of AI algorithms based on algorithms known as artificial neural networks[18]. These algorithms can now beat humans at pattern recognition tasks and at sophisticated games like poker or go[19]. But the hardware on which they run, such as graphical or tensor processing units limits their development outside large and particularly energy intensive data centers[20]. Therefore, developing hardware that is better suited to run current neural networks is an important challenge.

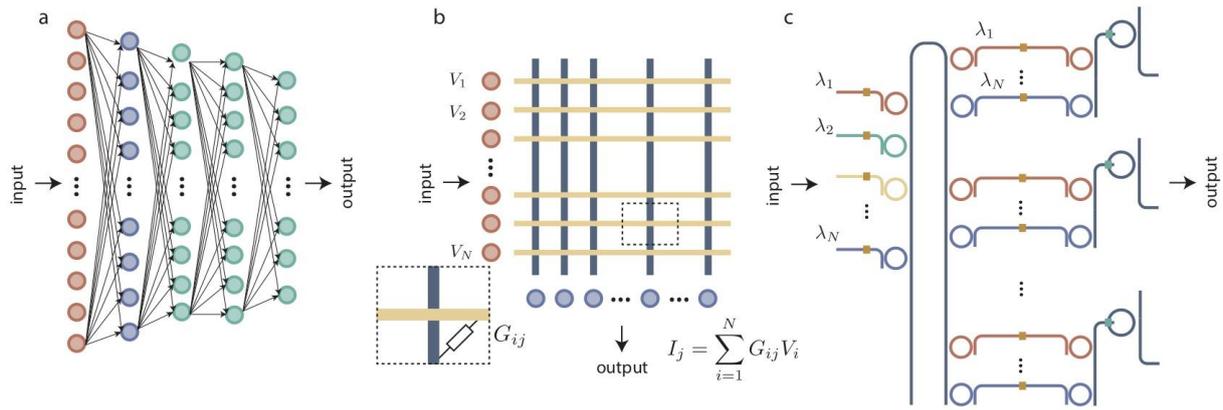

*Figure 1: Hardware for deep neural networks* (a) Deep neural network composed of layers of neurons (circles) connected by synapses (arrows). (b) Memristive crossbar array connecting two layers of neurons[21]. The inset represents a single memristor cell, vertically connecting a row and a column. The pre-synaptic CMOS neurons (in red) apply voltages to the rows. The output current Ij at each column is the sum of all input voltages Vi weighted by the memristor conductances Gij. An amplifier at each columns drives the post-synaptic CMOS neuron (in blue). (c) Optical neural network composed of circular resonators that couple different wavelength $\lambda_i$ inputs (in different colors) to a neuron (in gray)[22]. Synapses (orange squares) and neural activation functions (green squares) are implemented by phase change materials.

The highest accuracy neural networks today are called deep networks because they are composed of a large number of neuron layers, ranging from five to hundreds (Figure 1a). They typically take as input high-dimensional data, such as megapixel images, and reduce the dimension by outputting only a few classes, designating, for example, the content of the image (cat, dog, etc.). To achieve this, formal neurons and synapses in these networks are retaining only basic features of their biological counterparts. First, in most artificial neural networks, neurons are reduced to a mathematical function applied to their real valued inputs. This activation function should be non-linear, such as a sigmoid or a Rectified-Linear Unit (ReLu), in order to regulate the information transmitted to the next layer. Second, synapses are valves for the information, configuring how it flows in the network, guiding different inputs to the relevant output. These valves can transmit information positively as well as negatively, and are described by real valued synaptic weights. The input to a neuron in layer k+1 is the output of neurons in layer k multiplied by synaptic weights. This weighted sum operation, also called Multiply-And-Accumulate (MAC), is a key computation of artificial neural networks.

The weight values are initially chosen randomly, and the network needs to learn proper weight values before it can classify the inputs as desired. The most successful way to train networks today is through supervised learning. In this case, previously labeled examples are presented to the network. Its output is then compared to the known desired output, and an error is numerically computed. This error can then be backpropagated through the network to change each weight in proportion to its contribution to this error. Batch after batch of examples, the network can eventually converge to a point where the error is minimal, and where it can generalize when presented with examples that it has never seen during training.

Neuromorphic chips have the potential to accelerate both the inference phase (when a network is presented with an input and computes the output), and the training phase. A challenge is to build systems compatible with the hierarchical layered structure of the most powerful neural networks today, in which neuron outputs of one layer naturally feed the synaptic inputs of the next layer. As of today there are two main technologies for which scalable architectures have been proposed for mapping deep networks on chip using physics: hybrid CMOS/memristive and photonic systems.

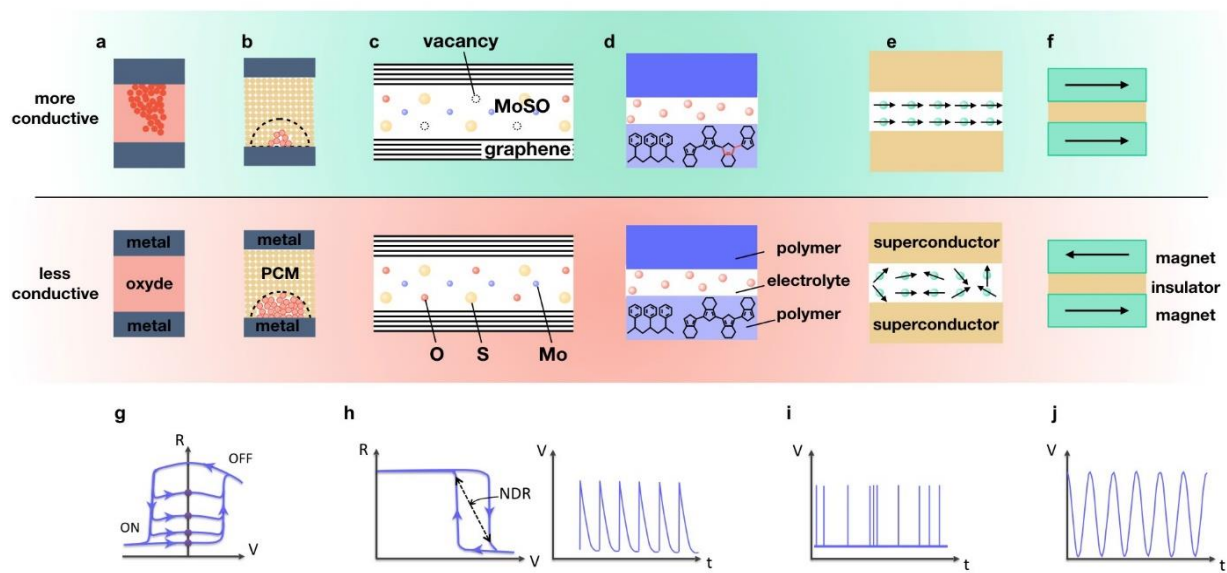

*Figure 2: Materials and physics for neuromorphic computing* (a-f) Different physical implementations of synapses. Top row in green corresponds to the highly conductive and the bottom row in red to the low conductive synaptic state. (a) A filamentary device (oxide or conductive bridge): the size of the filament between the top and bottom electrodes determines the resistance. (b) Chalcogenide-based

*phase change memory: the size of the amorphous region determines the resistance. (c) Van der Waals heterostructure graphene/MoSO/graphene[23]: the concentration of oxygen vacancies determines the resistance. (d) Organic electrochemical device[24]: conduction is assured by the positive ions in the electrolyte layer. A positive presynaptic voltage causes polymer reduction in the postsynaptic electrode, which absorbs some of the conducting ions and thus decreases the synaptic conductance. (e) Josephson junction with a magneticl barrier[25]: the magnetic order in the barrier determines the resistance. (f) Magnetic tunnel junction: the relative orientation of the magnetic layers determines the resistance. (g) Electrical resistance versus applied voltage of a non-volatile memristor, exhibiting multiple resistance states. (h) Electrical resistance versus applied voltage of a volatile memristive switch with negative differential resistance (NDR), Voltage spikes generated by such a device, (i) Voltage spikes generated by a resistively and capacitively shunted Josephson junction and (j) Time evolution of the electrical resistance of a superparamagnetic tunnel junction.*

In hybrid CMOS/memristive systems, neurons are made of analog or digital CMOS, and the information that flows through the network is electrical current. A synapse should therefore act like a valve for the current. This is exactly what a memristor, short for memory-resistor, is. Memristive devices, also called resistive switching devices, are nanoscale resistors with non-volatile analog conductance states tunable by an applied voltage (Fig. 2g). Such features can be obtained within a wide range of materials, and through different physical effects (Fig. 2a-d): electrical-field induced creation and control of nanoscale conductive filaments bridging two metallic electrodes separated by an insulating oxide such as tantalum or hafnium oxide[26], phase transitions leading to conductance changes in materials such as chalcogenides (structural phase transition, caused by Joule effect)[27] or strongly correlated electron systems[28] (electronic phase transition, caused by electric field), voltage-induced control of the ferroelectric configuration in the insulating barrier of ferroelectric tunnel junctions[29], and many other. When these memristors are arranged in a crossbar array configuration, they can be used to fully connect a layer of neurons to the next one[21,30] (see Figure 1b). The current going out of each of the bottom electrodes is indeed the weighted sum of the input voltages (applied at each row) by the conductances of the memristors in the column. Therefore, memristors directly implement the multiply-and-accumulate operation through Kirchhoff's laws (two memristors are necessary for one signed weight), an operation that is costly in terms of silicon area and energy consumption when achieved with CMOS. Purely resistive arrays have a limited size due to current sneak paths (i.e. current flowing from a row to a column through several memristors instead of one). This maximum size can reach few hundreds by few hundreds memristors, for devices with nonlinear current-voltage characteristics inherently limiting the impact of sneak paths[31]. Roads currently explored to achieving higher connectivity consist of assembling different crossbars together through CMOS-based buffers, either side by side[32], or on top of each other[33]; or alternatively, by adding a selector device (either a transistor or a volatile memristive switch) below each memristor. These advantages might not be limited to inference. It has been recently estimated through combined experiments and simulations that memristive systems could enable training neural networks with a hundred-fold gain in terms of energy consumption and speed compared to graphical processing cards[34]. Achieving learning with memristors, nevertheless, raises challenges addressed in the final section of this review.

Another neuromorphic computing approach scalable to deep networks has been proposed recently, based on photonic neural networks, which can be made with optical components only or mix optics with electronics using optoelectronic devices. Neurons can be implemented by optical resonators and synapses either by combining multiple interferometers or by modifying the transmission of optical waveguides with optically active phase change materials deposited on top [22,35,36] (see Figure 1c). Simple tasks have been demonstrated with these systems, such as vowel recognition[35]. The advantages of optics for computing are the possibility to convey wide amount of information in parallel on a single fiber or wave guide through massive wavelength multiplexing, and to build purely passive neural network with an extremely low energy consumption. On the other hand, the size of neurons and synapses might be difficult to shrink down below hundreds of micrometers, and the energy cost of converting the information to light and injecting it with power consuming lasers should be considered.

***Matching neuroscience-inspired concepts to hardware.*** The brain is much more complex than current AI algorithms. Biological neurons are more than non-linear functions. They spike, are leaky, feature memory, can be stochastic and can oscillate and synchronize (Figure 3a). They are spatially extended, with different functional compartments integrating signals coming from different areas[37,38].

Biological synapses are also more than analog weights. They are leaky memories, they have different time scales and different state parameters ruling their modifications. They can also be highly stochastic, as some synapses transmit only a fraction of the spikes that they receive[39] (box 2). Frequently ignored brain components play an important role and might be important to emulate: dendrites seem to be able to deploy extremely complex computations[40,41], while astrocytes (Figure 3a) seem to play a key role in neuron regulation[42]. All these properties can bring additional features to artificial neural networks, and are therefore interesting to implement and test.

All these ideas are especially promising if they can be implemented at low energy through the intrinsic properties of materials and related physical effects. This field of research, pioneered by Carver Mead, was originally focused on exploiting the exponential dependence of transistor leakage current on voltage [43–45]. In the last ten years, a wide range of other physical phenomena, depicted in Figure 2, have been used to mimic interesting properties of synapses and neurons.

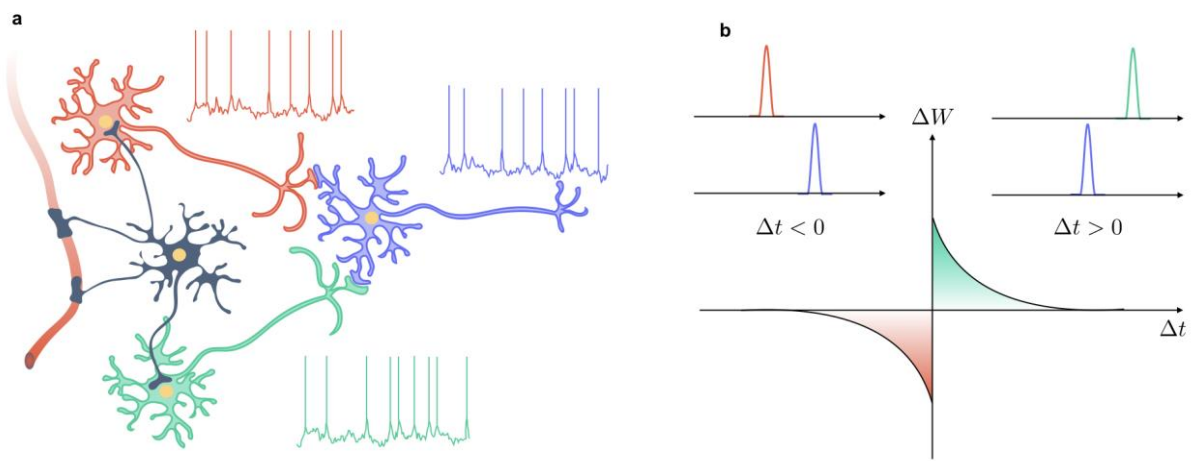

*Figure 3: Biologically-inspired neuromorphic computing (a) Two pre-synaptic neurons (red and green) connected to a post-synaptic neuron (purple). An astrocyte (grey) provides nutriments and energy from a capillary to neuron somas. All the neurons emit spike trains of action potential. (b) Spike-Time-Dependent-Plasticity. Here the pre-synaptic spike from the green neuron arrives shortly before the post-synaptic spike from the purple neuron (Δt >0), suggesting a correlation between these two events. The synapse connecting the two neurons is meaningful and its weight is increased (ΔW >0). On the contrary, the pre-synaptic spike from the red neuron arrives shortly after the post-synaptic spike. The weight of the synapse that connects them is decreased (ΔW <0).*

In particular, oxide electronics can imitate the multifunctionality of synapses and neurons. Biological synapses memorize meaningful information for long periods of time but rapidly forget unimportant data. Conductive bridge devices (Figure 2a) can emulate this dual long and short term memory nature of synapses[46,47]. Low amplitude pulses applied to the device trigger metallic filament growth between the electrodes. When the pulses are infrequent, the filament shrinks back, giving rise to a short-term, leaky memory. When the pulses are on the contrary frequently repeated, the filament does not have time to relax and grows until it strongly bridges the two electrodes, giving rise to long term memory. As we have seen, biological neurons emit spikes when their membrane potential exceeds a threshold. Materials exhibiting electronic phase transitions such as Mott insulators can emulate spiking neurons. When they are excited with series of incoming voltage spikes, an avalanche effect occurs and locally triggers the transition to a conductive state. If the input remains silent, the material slowly relaxes back to its original state. These two phases successively mimic neuron depolarization and repolarization during an action potential[48,49]. Volatile resistive switching devices exhibiting thermally-induced negative differential resistance can also be combined with additional capacitors, resistors or transistors to generate trains of spikes under a constant input voltage (Figure 2h)[48,50–53]. The spiking behavior is rich, leading to periodic spiking, chaotic spiking and bursting as in biological neurons[54,51,55]. In the brain, neurons influence their respective dynamics through their coupling via synapses, and can synchronize if synapses are strong[7]. The synchronization of neuron spikes enhances their efficiency for modifying synapses and triggers long term learning[56]. Memristive oxide synapses can control the dynamical coupling between spiking CMOS neurons, and lead to their synchronization[57].

Chalcogenide-based phase change memories (Figure 2b) are important devices for neuromorphic computing because of their technological maturity. They can implement non-volatile synapses for electronics and optics, as they exhibit memristive behavior[34,58] and can coat optical waveguides to tune the transmitted light[22]. The stochasticity of the phase transition between the amorphous and crystalline states has also been harnessed to implement stochastic nano-neurons whose membrane potential is represented by the phase configuration of the device[59].

2-D materials are composed of atomically thin layers, that can be gated and assembled in van der Waals heterostructures to generate a variety of behaviours, such as memristive behaviour through ionic motion[23,60] (Figure 2c). Resistive switching in 1D and 2D materials can be controlled through light, which provides the possibility to build bio-inspired optical sensors for neuromorphic chips.

In general, organic materials have the advantage of their compatibility with biology and low cost fabrication. They are flexible, multifunctional and often low-power. Electrochemistry has been for instance harnessed to implement low-voltage artificial synapses[24] and neurons[61] (Figure 2d). These devices, however, tend to suffer from slow operation due to the low mobility of carriers in organic materials, and to particularly high levels of variability.

In optics, non-linear dynamical phenomena can be harnessed for neuromorphic computing. Artificial neural networks used for AI today are mostly static and feedforward. On the contrary, biological neural networks are dynamic (their components evolve in time), and rely on recurrent loops. These features are useful to implement short term memory – the one we use to remember sequences of numbers or the beginning of a sentence[62]. The low attenuation of optical fibers has been leveraged to build recurrence loops. Delayed feedback optical oscillators can implement working memory[63]. Using an approach called reservoir computing, they can even perform classification and prediction tasks that require a dynamically stored memory[64–66]. Spatial light modulators are currently used to increase the number of neurons in these networks[67].

Flux quantization in superconductive Josephson junctions provides an interesting analogue to voltage spikes (Figure 2i). Rapid single flux quantum circuits can be adapted to emulate neuron spike emission and interconnection with extremely low power and dissipation[68,69]. Superconductive synapses have also been experimentally demonstrated. Magnetic cluster insertion in the barrier can for instance transform a Josephson Junction in an analog valve for the input current[25] (Figure 2e).

The cyclability of spintronic devices, together with their nanoscale dimensions and multifunctionality are particularly promising for neuromorphic computing[70]. Their stochastic switching and rich non-linear dynamics have been exploited to demonstrate different types of nano-neurons that exploit magnetization dynamics for computing[71,72] (Figure 2j).

Nano electro mechanical systems provide another type of dynamical system that can emulate neurons through their resonant or self-sustained oscillations[73]. A major interest for neuromorphic computing is their frequency compatible with audible sounds opening the route to on-chip voice recognition[74].

Among these different bio-inspired approaches, the most interesting are likely those that will allow crafting synapses and neurons with the same materials, and implementing learning algorithms through the physics of these devices.

## Developing physical neuromorphic systems in a lab

Most of the demonstrations that we have mentioned until now concern single or a limited number of devices. How is it possible to move from individual device physics to systems?

***Algorithms and physical computing substrates should be developed hand in hand.*** For neuromorphic computing, the materials, the physics that allows nanodevices to embody remarkable functions, and the bio-inspired computing models need to be thought and developed together. This is equally true for both artificial intelligence and neuroscience-inspired computing.

There are two main challenges at the device level regarding learning. The first one is precision. Learning through standard backpropagation, the most effective procedure for training deep AI networks, implies updating the weights after each batch of examples is presented, with variations that can be much smaller than 0.1% the value of the weight itself, due to vanishing gradient effects[75]. This is not an issue when weights are encoded in 32 bits floating point as is usually done on graphics cards, but more difficult to achieve with noisy nanodevices, whose effective weight values do not vary perfectly continuously[76]. The second challenge is achieving weight-independent weight variations with an applied voltage or current. Device non-linearities, which are often at the core of memristor or any artificial nano-synapse physics, will indeed typically prevent backpropagation to converge[34,76]. There is therefore today a considerable effort to produce devices with smooth and linear features. On the other hand, the learning mechanism intrinsically copes well with device-to-device variability, as discussed in the final section of this review.

In parallel to research work aiming at perfecting current devices, a considerable effort consists in adapting existing algorithms so that they can work despite the imperfections and unreliability of the physical substrate[32,58,77]. Recent success has been obtained in deep networks with binary weights instead of analog ones, which constitutes a major simplification for hardware implementations[78]. State of the art results have been obtained during the inference phase, however on-line learning still requires real values for the weights[79,80]. Physics-inspired Restricted Boltzmann Machine algorithms are more tolerant to memristor non-linearity during learning[81,82]. Other solutions work both at the algorithm and device level. For example, the stochasticity and non-linearity of memristive synapses can be alleviated by combining several devices together[58], or by using a linear capacitor as a weight for learning, but transferring regularly the weight value to a non-volatile memristor before it is forgotten[34].

On the other hand, the brain is able to learn with unreliable, and stochastic components. We should therefore be able to invent novel algorithms that allow learning with imperfect dynamical, noisy elementary nanodevices. In this direction, unsupervised learning is particularly interesting. One of the main challenges in AI today is indeed to develop algorithms that do not need millions of labeled examples to learn. A promising unsupervised learning algorithm is Spike Timing Dependent Plasticity (STDP)[83], a learning rule inspired by measurements in biology[84], where the synaptic weight of synapses is modified solely depending on the timing of spikes occurring on both sides of a synapse (see Figure 3b). Unlike error backpropagation, this learning strategy is spatially local and memristive devices can implement it naturally. A first strategy is to connect non-volatile memristors with neurons that display a particular shape of pulses[85,86]. Overlap occurring between voltage pulses on both sides of the synapse can cause the application of high voltages across a device and the modification of its conductance.

Another elegant strategy is to harness the volatility of some memristive devices to implement the learning rule directly through device physics[87,88]. A spike on one side of the synapse can for example briefly elevate the temperature of a device through Joule effect for a short period, causing any subsequent spike during this period to have an enhanced effect on device conductance. First demonstrations of pattern recognition through STDP with small ensembles of memristors have been recently achieved[89,90], and with larger ensembles on phase change memories[82]. Currently, the biggest challenge for scaling STDP-based systems might not come from devices and materials, but from the limitations of the STDP rule itself. Newer theories are needed to extend the cognitive capabilities of STDP to complex multilayer systems, and the availability of memristive devices is one of the core motivations for such research[91,92].

We can even imagine algorithms that exploit nanodevices defects or dynamics for information processing. For example, noise can be harnessed for computing in many different ways including stochastic resonance[93], energy minima exploration[72,94] and diffusion[95]. As brain networks exhibit criticality, phase transitions, synchronization: can these physical effects be harnessed in-materio for computing and learning? These questions are currently explored by many groups, theoretically and experimentally.

***Small-scale "toy" systems.*** Building large scale systems often requires access to a CMOS foundry and to industrial partners. However, academic laboratories can make important contributions to system-level developments by developing "toy" physical neuromorphic systems. The idea is the same as well known "toy models" in physics: develop small physical neuromorphic systems to test some hypothesis and draw some conclusions. Recent experiments in the field have shown that realizing such systems can bring important insights not necessarily identified by theoretical investigations and single device studies.

We are for instance developing neural networks composed of spintronic nano-oscillators as neurons. In our first experiment[71], we wanted to check if a spintronic oscillator could be used as a neuron to compute. For this purpose, we used the approach of reservoir computing, which makes it possible to emulate a full neural network with a single oscillator multiplexed in time[64] (Figure 4a). We learned from this experiment that oscillator noise is a severe issue for reservoir computing. In this approach, the oscillator output is indeed sampled and the resulting data points are linearly combined to form the output. To obtain reliable pattern recognition, the oscillator should therefore not drift, and it should feature a high signal to noise ratio. This is problematic at the nanoscale, as fluctuations and drift effects often arise due to the small volume of devices[70]. However, we found an exploitable range of conditions where spintronic nano-oscillators have enough stability and signal to noise ratio to obtain performance at the state of the art of macroscopic oscillators and even software[71]. This is due to the exceptional cyclability of spintronic nanodevices and the possibility to finely tune their dynamical response through current and fields.

Computing with noisier oscillators is possible but requires engineering redundancy at the neuron level so that ensemble average can suppress the effects of noise. We tested this strategy with another kind of oscillators, superparamagnetic tunnel junctions[72,93], which are actually noisy to the point that they behave in an entirely stochastic fashion (Figure 3j). We took inspiration from a neuroscience theory called population coding to implement the computation of various non-linear functions[96].

After our first experiment with a single spintronic nano-oscillator, we moved on to our next neuromorphic system: a small neural network formed of four coupled spintronic nano-oscillators[97] (see Figure 4 (b)). We wanted to investigate if it was possible to achieve pattern recognition by synchronizing some of the oscillators to incoming alternating inputs, encoding, in their frequency, the signal to classify. The experiment worked, and was able to classify spoken vowels[97]. However, it required the oscillators to be highly tunable in frequency. Nano-oscillators indeed feature an unavoidable variability in their properties. This variability in the basic components of a neural network can be compensated during training, which was achieved, in our small neural network, by exploiting the high tunability of the frequency of spintronic oscillators with the injected direct current (reaching tens of percent of the base frequency).

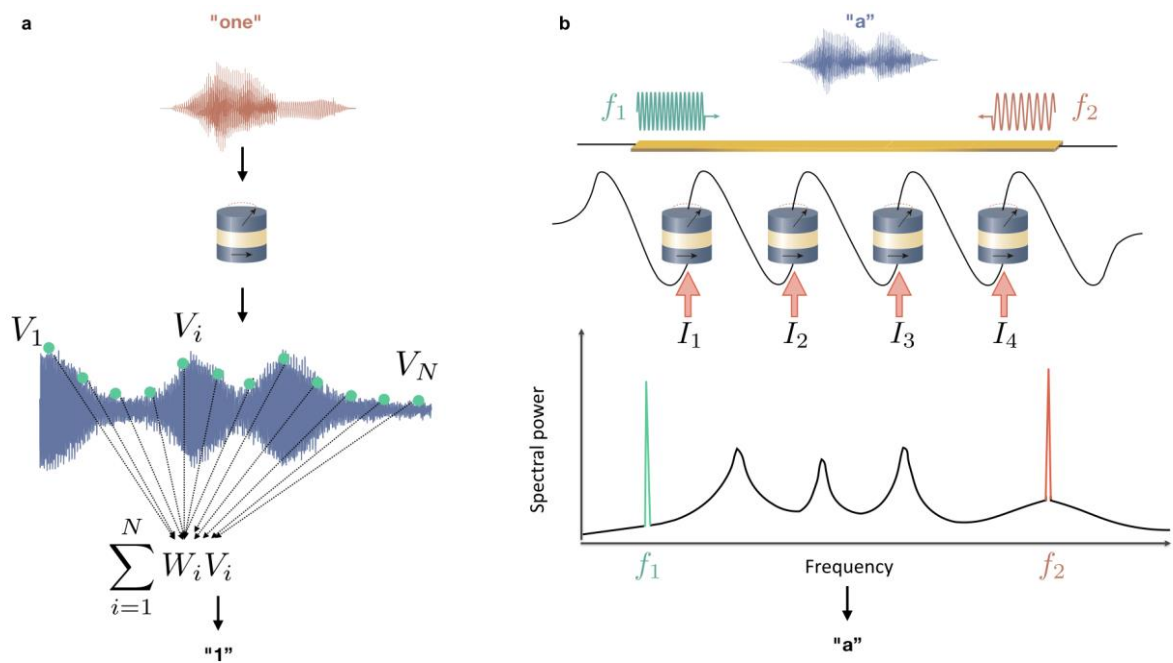

*Figure 2 : Toy neuromorphic systems with spintronics (a) Reservoir computing with a single time-multiplexed spintronic nano-oscillator. The spoken digit (here "one") to recognize is preprocessed into an input injected to the nano-oscillator, emulating different neurons at different time steps. The output is a linear combination of the nano-oscillator emitted voltage at different time steps. (b) Pattern recognition from synchronization patters of four coupled spintronic nano-oscillators. The spoken vowel to recognize is preprocessed into two input RF signals f1 and f2 injected to the network by a common waveguide. The nano-oscillators are electrically coupled to each other and capable of synchronizing to the input signals. Their frequencies are tuned by injection of individual dc-currents I1 to I4. Each synchronization pattern (here oscillator 4 is synchronized to input frequency f2) corresponds to one learned vowel (here "a").*

***Developing large scale systems.*** The final goal of neuromorphic computing is to build large scale systems comprising millions of components. Therefore, care should be taken to study materials and physics that are scalable. Scaling physics-rich neuromorphic systems raises several major challenges. First, densely and efficiently interconnecting nanosynapses and nanoneurons is a major challenge of neuromorphic computing. Here, two strategies are possible. It is possible to step away from the bioinspiration and to utilize CMOS circuits for interconnection. Then, conventional electrical engineering methods can be used to route signals between neurons and synapses, in particular multiplexing techniques and routers to limit the amount of interconnections. This means that the difficulty of achieving massive connection in electronics can be compensated by the availability of fast digital circuits to achieve routing. The danger, in this situation, is of course to lose the advantages of using emerging nanotechnology in the first place, due to the CMOS overhead in terms of area and energy consumption. For this purpose, the CMOS circuits need to exploit as well as possible the physical properties of the nanotechnology, through a radical co-design approach. The other strategy is to harness physics, materials and nanotechnology to densely connect the neural network at the system level, and achieve densities above one thousand synapses per neuron, approaching the brain interconnection level. A pioneering idea in this direction are the 3-D crossnets[33,98], in which memristor crossbar arrays are stacked on top of each other. Other approaches include wireless communication between neurons and synapses, using optics[64,67] and microwaves[97], as well as self-assembly in 2-D and 3-D[99].

The second challenge to realize large scale neuromorphic systems with physics is to deal with nanodevice variability. Training the network on-chip is the best way to compensate these device variations. Indeed, through learning, each component (synapses, and sometimes neurons) can be tuned to deliver the desired output taking into account the specificities of the devices it is connected to[34,82,94]. Another exciting lead is to implement systems where the device variability is seen as an asset, not an issue. The CMOS-based Braindrop system[100] for example fully embraces device mismatch, as transistor variability is used to provide a collection of circuits with a wide range of behaviors, which provide basis functions for building complex functionalities. On the other hand, the variability of memristive devices can be leveraged to implement algorithms such as Markov Chain Monte Carlo that require a massive amount of random numbers, which the nanodevices naturally provide through their cycle-to-cycle variability[94].

# Conclusion and perspectives

We have highlighted in this review the relevance of physics for neuromorphic computing. It is a striving field, in which many approaches are currently tested, which all have their advantages and disadvantages (Table 1), and the question of what will be the materials and physical principles used in future neural network hardware is still open-ended.

|  | CMOS synapses and neurons | Resistive switching synapses with CMOS neurons | Photonic synapses and neurons | Spintronic synapses and neurons | Superconductive synapses and neurons |
|---|---|---|---|---|---|
| Connections | wires | wires | Light | microwaves | Wires or microwaves |
| Min neuron lateral size | 10 μm | 10 μm | 100 μm | 10 nm | 20 nm |
| Min synapse lateral size | 10 μm | 10 nm | 1 μm | 10 nm | 20 nm |
| Advantages | commercial | Nanoscale synapse, technology-ready | Wavelength multiplexing, can be totally passive (zero energy consumption) | Nanoscale synapses and neurons, almost commercial technology | Low energy consumption beside cryogenic requirements, all identical spikes |
| Disadvantages | Size of neurons and synapses, no in-memory computing | Size of neurons, complex wiring | Size of neurons and synapses, dissipation due to lasers | Scalability to be demonstrated | Scalability to be demonstrated |
| Chips | Inference | Inference coming soon | no | no | no |

*Table 1: Comparison between the different approaches to neuromorphic computing*

Qubits and quantum oscillators are conspicuous absents in Table 1. Until recently, the fields of quantum computing and neuromorphic computing were evolving in parallel. However, cross-fertilization between the two domains has started, and is likely to bring remarkable results. Building physical neural networks of quantum neurons and synapses opens possibilities to make use of quantum superposition and entanglement to process information in parallel and of high dimensional state space to implement deep neural networks, while taking advantage of neuromorphic computing to deal with noise and device variations[101–103].

As we have pointed out in this review, neuromorphic computing with emerging nanodevices, physics and materials is a fast evolving field. Recently, different systems using more than a million memristors have been reported[104,105]. These first fully integrated systems are functional and report reasonable raw accuracy on practical tasks, nevertheless remaining significantly lower than the standards of software AI. However, it should be understood that due to the extremely long delays associated with fabricating and testing functional electronic systems incorporating emerging technologies, these chips were designed a significant time ago, and therefore do not incorporate the latest progresses of the field. We should expect exciting new results coming from both academic and industry labs soon. Currently, one of the most optimized fully integrated inference systems reported[106] achieves 94.4% accuracy on the canonical MNIST task of handwritten digit recognition. By contrast, to this day, achieving large functional systems capable of learning remains a challenge. Most works use hybrid approaches where a part of the system is simulated in software[34,94]. Ref.[105], nevertheless, is a fully hardware system implementing STDP learning with 1.4 Million synapses that, despite some limitations inherent to early generation developments, shows the way for future brain-inspired hardware exploiting physics.

In parallel to the development of these new physical neural networks, solutions based on conventional digital CMOS are improving at high pace, due to the demand for fast application-oriented AI processors. Recent GPUs are co-designed to minimize data motion and reduce bit precision, thus achieving extremely low power consumption for specific neural network types and applications[107]. Evolving even faster is the field of artificial intelligence, in which novel algorithms and learning techniques are proposed at an extremely fast pace and immediately adopted by the community. A challenge and impetus for the physics based-neuromorphic computing is to keep up with the advances on these two fronts: conventional CMOS chips, and AI algorithms. The first solution is to build radically interdisciplinary teams that develop hand in hand the algorithms and the hardware. The second solution is to identify the challenges that current CMOS and AI algorithms will not be able to give an answer to in the long term, and offer new solutions to these issues based on novel physics and materials. This is what we have tried to highlight in this review, especially stressing the importance to take inspiration from the blooming field of Neuroscience, as well as Artificial Intelligence.

***Box 1: Brain versus Computer – comparing the energy efficiency***

Here, we compare the brain[108] and a supercomputer training the deep neural network BERT[109,110].

The brain performs and learns multiple tasks in parallel, such as natural language processing, vision, auditory processing, synthesis, social interactions, sensory-motor coordination, basic mathematics and so on. The neural network BERT is state-of-the art in natural language processing.

With an energy budget of 1 000 kW.h (3.6 GJ), the brain can operate during six years. BERT, with the same energy budget, can only be trained during 80 hours using 64 Tesla V100 GPUs.

The brain consumption of oxygen and glucose represents 20% of the whole body power consumption, amounting to 20 W for an adult. In the supercomputer, the power consumption comes mostly from the CPU, the GPUs and the DRAM.

The brain is composed of about $10^{15}$ synapses and $10^{11}$ neurons. The neural network BERT has $10^8$ tunable synaptic weights.

It is difficult to establish a fair comparison of the energy consumption of the elementary operations in the brain and in a supercomputer. The energy consumption of spikes and synaptic events in the brain is known, and comes mostly from moving $Na^+$ and $K^+$ ions across cell membranes. It amounts to $3.8 \cdot 10^8$ Adenosine Triphosphate (ATP) molecules or 60 pJ for a neuron spike, and $1.6 \cdot 10^5$ ATP or 26 fJ for a synaptic event. Such calculation is more difficult for a supercomputer which does not contain physical synapses and neurons, and delocalizes the operations in different parts of the processors. What we can say is that training BERT takes $10^6$ steps, each composed of $10^5$ words and for which all weights are modified. Therefore, the energy per weight modification can be estimated to 100 pJ.

The figures given above highlight the striking differences in the overall energy consumption of the brain and a supercomputer. It is much more difficult to compare the energy consumption of the brain and a supercomputer for the exact same task. Contrary to artificial neural networks, the brain never stops from learning and there is no separation between the costs of inference and learning. Intense task focus on a task in the brain increases the local consumption by less than 10 %. Being in a coma reduces the consumption by 50 %, sleeping by 10 to 30 %. For the supercomputer, we took the case where the neural network is trained on a single task. However, the network architecture (number of layers, of neurons and parameters) was previously optimized. The cost of learning from scratch a new dataset involves tuning *hyperparameters* and architecture search, which was not included in our calculations, can lead to up to a 1,000 fold increase in the total power consumption.

*Box 2: Stochasticity in the brain*

It has been observed that in the brain:

- Ions channels open and close stochastically: the resulting current is telegraphic signal. Because there are typically 100 ion channels simultaneously, this causes $\sqrt{100}/100$ = 10% fluctuations[111]. This degree of fluctuation is considerable compared to the reliability of conventional microelectronic circuits.
- Synaptic release is probabilistic[112]. The release probability is highly synapse-dependent and varies with parameters such as the spiking rate and temperature, with typical values around 25%. This mechanism differs fundamentally from the deterministic switching of CMOS gates.

An interpretation of this stochasticity is that the brain may maximize not the information transfer itself but the ratio between information transfer and energy[39,113].

**Acknowledgements**

This work was supported as part of the Q-MEEN-C, an Energy Frontier Research Center funded by the U.S. Department of Energy (DOE), Office of Science, Basic Energy Sciences (BES), under Award DE-SC0019273 (reservoir computing with feedback) and by the European Research Council ERC under Grant bioSPINspired (682955) and NANOINFER (715872).

**Author contributions**

All authors wrote the review.

**Author information**

The authors declare no competing financial interests. Correspondence and requests for materials should be addressed to J.G. (julie.grollier@cnrs-thales.fr) and D.Q. (damien.querlioz@c2n.upsaclay.fr).